\documentclass[aps]{revtex4}  
\usepackage{graphicx}  
\usepackage{dcolumn}   
\usepackage{bm}        
\usepackage{amssymb}   
\def\MET{{\mbox{$E\kern-0.57em\raise0.19ex\hbox{/}_{T}$}}}
\def\met{{\mbox{$E\kern-0.57em\raise0.19ex\hbox{/}_{T}$}}}
\def\DZ{D\O\ }
\def\DZero{D\O\ }
\def\Dzero{D\O\ }
\def\ipb{~pb$^{-1}$}
\def\ifb{~fb$^{-1}$}
\def\pp{$p\bar{p}$}

\def\tt{$t\bar{t}$}
\def\WH{$WH\rightarrow \ell\nu b\bar{b}$}
\def\lmet{$WH\rightarrow \ell\kern-0.45em\raise0.19ex\hbox{/} \nu b\bar{b}$}
\def\ZH{$ZH\rightarrow \nu\bar{\nu} b\bar{b}$}
\def\ZHll{$ZH\rightarrow \ell^+ \ell^- b\bar{b}$}
\def\zhe{$ZH\rightarrow e^+ e^- b\bar{b}$}
\def\zhm{$ZH\rightarrow \mu^+ \mu^-  b\bar{b}$}
\def\pwzh{$p\bar{p}\rightarrow W/ZH \rightarrow \ell \nu b\bar{b}/ \nu\bar{\nu} b\bar{b} / \ell^+\ell^- b\bar{b}$}
\def\pwww{$p\bar{p}\rightarrow WH \rightarrow WW^{+} W^{-}$}
\def\www{$WH \rightarrow WW^{+} W^{-}$}
\def\phww{$p\bar{p}\rightarrow H \rightarrow W^{+} W^{-}$}
\def\hww{$H\rightarrow W^+ W^-$}
\def\hbb{$H\rightarrow b\bar{b}$}

\def\tevE{$sqrt{s}=1.96$~TeV}

\begin{document}
\rightline{CDF Note 8384}
\rightline{\DZ Note 5227}
\vskip0.5in

\title{Combined \DZ and CDF Upper Limits on Standard-Model Higgs-Boson Production}

\author{The TEVNPH Working Group\footnote{The Tevatron
New-Phenomena and Higgs working group can be contacted at
TEVNPHWG@fnal.gov. More information can be found at http://tevnphwg.fnal.gov/.}}
\affiliation{\vskip0.1in for the \DZ
 and CDF Collaborations }
\begin{abstract}
We combine results of CDF and \DZ~searches for a Standard-Model Higgs
boson (H) in data from \pp~collisions at the Fermilab Tevatron with
$\sqrt{s}=1.96$~TeV. With 260-950\ipb~collected at \DZ,and
360-1000\ipb~collected at CDF, the 95\% CL upper limits are a factor
of 10.4(3.8) higher than the expected cross section for
$m_{H}=$115(160)~GeV/c$^2$. This result extends significantly the
individual limits of each experiment.
\end{abstract}

\maketitle

\section{Introduction} 
Because the mechanism for electroweak symmetry breaking has yet to be
confirmed, the search for a Standard-Model (SM) Higgs boson represents
a significant issue in Fermilab's Tevatron physics program. Both \Dzero and
CDF have recently reported searches for Higgs bosons that combined
different final states and production modes\cite{DZhiggs,CDFhiggs}.

In this note, we combine results of all such searches from CDF and
\DZ~for \pp~collisions at~\tevE. These searches are for SM Higgs
bosons produced in association with vector bosons (\pwzh~or \pwww) or
singly through gluon-gluon fusion (\phww). The results are for data
corresponding to integrated luminosities ranging from 360-1000\ipb~at
CDF and 260-950\ipb~at \DZ. The searches are separated into sixteen
final states, referred to as analyses in the following. To simplify
combination of signals, the analyses are separated into sixteen
mutually exclusive final states. Selection procedures for each
analysis are detailed in each of the experiment's
reports\cite{DZhiggs,CDFhiggs}, and briefly described below.

\section{Acceptance, Backgrounds, and Luminosity}  

Selections are similar for the corresponding CDF and \DZ analyses.
For the case of \WH, an isolated lepton (electron or muon) and at
least two jets are required, with one or more jets tagged as
originating from $b$-quarks.  Two orthogonal tagging criteria are
defined, one being an exclusive single-tag (ST) and the other a
double-tag(DT) selection. These events must also display a significant
imbalance of momentum in the plane transverse to the beam axis
(referred to as missing energy or \met).  Events with additional
isolated leptons are vetoed.  For the \ZH~analyses, the selection is
similar, except all events with isolated leptons are vetoed and
stronger multijet background suppression techniques are applied. As
there is a sizable amount of \WH~signal that can mimic the \ZH~final
state when the lepton is undetected, the \DZ~analyses include this as
a separate search, referred to as \lmet.  CDF includes this as part of
the acceptance of the \ZH~search.  In the \WH~and \ZH~analyses, the
final variable used for setting a cross section limit is the dijet
invariant mass. The \ZHll~analyses require two isolated leptons and
two jets, wherein the CDF analysis requires at least one of the jets
to be $b$-tagged while the \DZ~analysis uses only events with two
$b$-tags. For the \DZ analysis the dijet invariant mass is used for
setting limits while CDF uses the output of a 2-dimensional Neural
Network to discriminate between signal and background.  For the
\hww~analyses, a large \met~and two opposite-signed, isolated leptons
(electrons or muons) are selected, defining three final states
($e^+e^-$, $e^\pm \mu^\mp$, and $\mu^+\mu^-$).  The presence of
neutrinos in the final state prevents the direct reconstruction of the
Higgs mass, and the final variable used is the difference in $\varphi$
between the two final-state leptons. The \DZ experiment also
contributes three \www~analyses, where the associated $W$ boson and
the same-charged $W$ boson from the Higgs decay semi-leptonically,
thereby defining six final states containing all decays of the third
$W$ boson (of opposite charge). In this case of this analysis, the
final variable is a likelihood discriminant formed from several
topological variables.

All Higgs signals are simulated using \textsc{PYTHIA}
v6.202\cite{pythia}, using \textsc{CTEQ5L}\cite{cteq} leading order
parton distribution functions. The signal cross sections are
normalized to next-to-next-to-leading order
calculations\cite{nnlo1,nnlo2}, and branching ratios from
\textsc{HDECAY}\cite{hdecay}. For both CDF and \DZ, events from
multijet (instrumental) backgrounds (``QCD production'') are measured
in data, but with different methods. For CDF, inherent backgrounds
from other SM processes were generated using \textsc{PYTHIA},
\textsc{ALPGEN}\cite{alpgen}, and \textsc{HERWIG}\cite{herwig}
programs. For \DZ, inherent backgrounds were generated using
\textsc{PYTHIA}, \textsc{ALPGEN}, and \textsc{COMPHEP}\cite{comphep},
with \textsc{PYTHIA} providing parton-showering and hadronization for
all the generators.  Background processes were normalized using either
experimental data or next-to-leading order calculations from
\textsc{MCFM}\cite{mcfm}.

Values of integrated luminosity, and expected signal, expected
background, and observed events are given in Table~\ref{tab:cdfacc}
for CDF analyses and Tables~\ref{tab:dzacc1}-\ref{tab:dzacc3} for \DZ
analyses.  The numbers of events are integrated in the range $0\leq
m_{jj} \leq$200~GeV/c$^2$ for \hbb~analyses, $0\leq
\Delta\varphi(\ell_1, \ell_2) \leq\pi$ for \hww~analyses, and $0\leq
Discrim \leq 1$ for the \www~analyses. The tables also include the
value of the Higgs mass for which each set of numbers is derived.

\begin{table}
\caption{\label{tab:cdfacc}The luminosity, expected signal, expected
background, and observed data for the CDF analyses.  Also included is
the Higgs mass for which each set of numbers are derived. For the \ZH~analysis
the signal events are given separately for $WH$ and $ZH$ events. The numbers
of expected events are determined for $0\leq m_{jj} \leq$200~GeV/c$^2$
for \hbb~analyses, and $0\leq \Delta\varphi(\ell_1, \ell_2) \leq\pi$ for
\hww~analyses.}
\begin{ruledtabular}
\begin{tabular}{lcccc}
\\
&$WH\rightarrow \ell\nu b\bar{b}$ & $ZH\rightarrow \nu\bar{\nu} b\bar{b}$ & 
$ZH\rightarrow \ell^+\ell^- b\bar{b}$ &  $H\rightarrow W^+ W^- \rightarrow \ell^\pm\nu \ell^\mp\nu$ \\ 
&  DT(ST) &  DT(ST) & & \\\hline 
Luminosity (\ipb)                & 1000 & 1000 & 1000 & 360\\ 
Expected Signal (evts)     & 0.50 (1.47) & 0.25+0.22 (0.77+0.64)& 0.64 & 0.84\\ 
Expected Background (evts)& 40.2 (366.3) & 19.6 (309.9)& 103.7 & 13.8\\ 
Data (evts)                               & 36 (390) & 24 (333)& 104 & 18 \\ 
$m_{H}$ (GeV/c$^2$)           & 115 & 115 & 120 & 160 \\
Reference       & \cite{cdfWH} & \cite{cdfZH}& \cite{cdfZHll} & \cite{cdfHWW} \\
\end{tabular}
\end{ruledtabular}
\end{table}

\begin{table}
\caption{\label{tab:dzacc1}The luminosity, expected signal, expected
background, and observed data for the \DZ \hbb~analyses.  The number
of expected signal, background, and data are determined for
$0\leq m_{jj} \leq$200~GeV/c$^2$.}
\begin{ruledtabular}
\begin{tabular}{lcccccccc}
\\
&$WH\rightarrow e\nu b\bar{b}$ & $WH\rightarrow \mu\nu b\bar{b}$  & \lmet  & $ZH\rightarrow \nu\bar{\nu} b\bar{b}$ & $ZH\rightarrow \ell^+\ell^- b\bar{b}$ \\ 
& DT(ST) & DT(ST) &DT(ST) &  DT(ST) & \\\hline
Luminosity (\ipb)         & 371 & 385 & 260& 260& 320-389\\ 
Expected Signal (evt)     & 0.19 (0.22) & 0.11 (0.12)& 0.18 (0.20)&0.24 (0.26) & 0.11\\ 
Expected Background (evts)& 14.6 (58.3) & 10.2 (46.8)& 25.2 (87.3)&25.2 (87.3) & 11.0\\ 
Data (evts)               & 14 (57) & 8 (48)& 23 (98)&23 (98) & 14\\ 
$m_{H}$ (GeV/c$^2$)       & 115 & 115 & 115 &115  & 115\\
Reference       & \cite{dzWHl} & \cite{dzWHl}& \cite{dzZHv} & \cite{dzZHv} & \cite{dzZHll} \\
\end{tabular}
\end{ruledtabular}
\end{table}

\begin{table}
\caption{\label{tab:dzacc3}The luminosity, expected signal, expected
background, and observed data for the \DZ \www~analyses.  The number of expected signal, background, and data are
determined for $0\leq \Delta\varphi(\ell_1, \ell_2) \leq\pi$.}
\begin{ruledtabular}
\begin{tabular}{lcccccc}
\\
&$WW^+ W^- \rightarrow e^\pm\nu e^\pm\nu$ &$WW^+ W^- \rightarrow e^\pm\nu \mu^\pm\nu$ &$WW^+ W^- \rightarrow \mu^\pm\nu \mu^\pm\nu$ \\\hline 
Luminosity (\ipb)         & 384 & 368 & 363\\ 
Expected Signal (evts)    & 0.043& 0.101& 0.066\\ 
Expected Background (evts)& 15.4& 7.0& 12.5\\ 
Data (evts)               & 15& 7& 12\\ 
$m_{H}$ (GeV/c$^2$)       & 155 & 155 & 155 \\
Reference       & \cite{dzWWW} & \cite{dzWWW}& \cite{dzWWW} \\
\end{tabular}
\end{ruledtabular}
\end{table}

\begin{table}
\caption{\label{tab:dzacc2}The luminosity, expected signal, expected
background, and observed data for the \DZ \hww~analyses.  The number
of expected signal, background, and data are determined for $0\leq
\Delta\varphi(\ell_1, \ell_2) \leq\pi$.}
\begin{ruledtabular}
\begin{tabular}{lcccccc}
\\
&$H\rightarrow W^+ W^- \rightarrow e^+\nu e^-\nu$ &$H\rightarrow W^+ W^- \rightarrow e^\pm\nu \mu^\mp\nu$ &$H\rightarrow W^+ W^- \rightarrow \mu^+\nu \mu^-\nu$ \\\hline 
Luminosity (\ipb)         & 950 & 950 & 930\\ 
Expected Signal (evts)    & 0.64 & 1.50& 0.54\\ 
Expected Background (evts)& 11.4 & 28.1& 10.5\\ 
Data (evts)               & 11 & 18& 10\\ 
$m_{H}$ (GeV/c$^2$)       & 160 & 160 & 160 \\
Reference       & \cite{dzHWWee} & \cite{dzHWWee}& \cite{dzHWWmm} \\
\end{tabular}
\end{ruledtabular}
\end{table}

\section{Combination Procedures} 

To gain confidence that the final result does not depend on the
details of the statistical formulation, we combine all separate
results using both a Bayesian and a Frequentist approach.  In both
methods, distributions in the final variables are binned according to
their experimental resolution, rather than as single integrated
values. Systematic uncertainties enter as uncertainties on the
expected number of signal and background events in each analysis.
Both methods use likelihood calculations based upon Poisson
probabilities. 

\subsection{Frequentist Method}

The Frequentist technique relies on the $CL_s$ method, using a
log-likelihood ratio (LLR) as test statistic\cite{DZhiggs}, as given
by:

\begin{equation}
LLR_n = 2\sum_{i=1}^{N}\left(s_i - n_i \textrm{Log}(1+s_i/b_i)\right)
\end{equation}

\noindent where $n$ denotes the hypothesis being tested
(\textit{e.g.}, background-only or observed data) and the sum runs
over the number of bins (and/or analyses) being combined. The value of
$CL_s$ is then defined as the normalization of the signal+background
hypothesis ($CL_{s+b}$) by the background-only hypothesis ($CL_{b}$).
This construction reduces the ambiguity of ``unphysical'' results
({\it e.g.}, negative cross section limits) and separates properties
of the estimator from that of the quantity being probed.

\subsection{Bayesian Method}

Because there is no information on the production cross section for
the Higgs, the Bayesian technique\cite{CDFhiggs} assigns a flat prior
for the total number of Higgs events. For a given Higgs mass, the
combined likelihood is a product of the likelihoods in the individual
channels, each of which is a product over histogram bins:

\begin{equation}
{\cal{L}}(R,\vec s, \vec b |\vec n) = \prod_{i=1}^{N_C}\prod_{j=1}^{Nbins} \mu_{ij}^{n_{ij}} e^{-\mu_{ij}}/n_{ij}!
\end{equation}

\noindent where the first product is over the number of channels
($N_C$), and the second product is over histogram bins containing
$n_{ij}$ observed events, either in dijet mass for $WH$ and $ZH$, in
$\Delta
 \varphi$ of two leptons for \hww, or in the likelihood
discriminant for \www~events. The parameters that contribute to the
expected bin contents are $\mu_{ij} =R \times s_{ij} + b_{ij}$ for the
channel i and the histogram bin j. The posterior density function is
then integrated over all parameters (including correlations) except
for $R$, and a 95\% credibility level upper limit on $R$ is estimated
by calculating the value of $R$ that corresponds to 95\% of the area
of the resulting distribution.

\subsection{Systematic Uncertainties} 

Systematic uncertainties on background are generally several times
larger than the expected signal, and are therefor important in the
limit calculation. Each systematic uncertainty is folded into the
signal and background expectations assuming Gaussian distributions.
The Gaussian values are sampled once for each Poisson Monte Carlo (MC)
trial (pseudo-experiment). Correlations among systematic sources are
carried through in the calculation. Systematic uncertainties differ
between experiments and analyses, both for signal and background
sources.  Detailed discussions of these issues can be found in the
individual analysis notes\cite{DZhiggs,CDFhiggs}.  Here we will
consider only the largest contributions and correlations among and
within the two experiments.

\subsubsection{Correlated Systematics}

The uncertainty on the measurement of the integrated luminosity is 6\%
(CDF) and 6.5\% (\DZ).  Of this value, 4\% arises from the uncertainty
on the inelastic \pp~scattering cross section, which is correlated
between CDF and \DZ. The uncertainty on the production rates for
top-quark processes (\tt~and single-top) and electroweak processes
($WW$, $WZ$, and $ZZ$) are also taken as correlated between the two
experiments. As the methods of measuring the multijet (QCD)
backgrounds differ between CDF and \DZ, there is no
correlation assumed for this uncertainty.

\subsubsection{CDF Systematics}

The dominant systematic uncertainties for the CDF analyses are shown
in Table~\ref{tab:cdfsyst1}. For \hbb, the largest
uncertainties on signal arise from the $b$-tagging scale factor
(5.3-16\%), jet energy scale (1-20\%), and MC modeling (2-10\%). For
\hww, the largest contributing uncertainty comes from
MC modeling (5\%).  For inherent backgrounds, the uncertainties on the
expected rates range from 11-40\% (depending on background). Because
the largest background contributions are measured using data, these
uncertainties are treated as uncorrelated for the \hbb~channels. For
the \hww~channel, the luminosity uncertainty is taken to be correlated
between signal and background. The differences between treating the
remaining uncertainties to be correlated or uncorrelated is less than
5\%.

\begin{table}
\caption{\label{tab:cdfsyst1}The breakdown of systematic uncertainties
for each individual CDF analysis. All positive-signed uncertainties
within a group are considered 100\% correlated across channels.
Values with negative signs are considered uncorrelated. }
\begin{ruledtabular}
\begin{tabular}{lcccccc}
\\
Source &$WH\rightarrow \ell \nu b\bar{b}$ ST & $WH\rightarrow \ell \nu b\bar{b}$ DT & $ZH\rightarrow \nu\bar{\nu} b\bar{b}$ ST & $ZH\rightarrow \nu \bar{\nu} b\bar{b}$ DT & $ZH\rightarrow \ell^+ \ell^- b\bar{b}$ & $H\rightarrow W^+ W^-$\\\hline
Luminosity (\%)& 6.0 & 6.0 & 6.0 & 6.0& 6.0 & 6.0 \\
$b$-Tag Scale Factor (\%)& 5.3 & 16.0 & 8.0 & 16.0 & 8.0 & n/a\\
Lepton Identification (\%)& 2.0 & 2.0 & 2.0 & 2.0 & 1.4 & 3.0\\
Jet Energy Scale (\%)& 3.0 &3.0 & 6.0& (1.0-20.0) & (1.6-20.0) & 1.0\\
I(S)R+PDF (\%)& 4.0 & 10.0 & 4.0 & 5.0 & 2.0 & 5.0 \\
Trigger (\%) & 0.0 & 0.0 & 3.0 & 3.0 & 0.0 &0.0 \\
$Z+h.f.$ Shape   (\%)& n/a & n/a  & n/a & n/a& -20& n/a \\ \hline
\\
Backgrounds\\\hline
W/Z+HF(I) (\%) & 33.0 & 34.0 & 12.0 & 12.0 & 40.0 & n/a \\
W+HF(II) (\%) & 0 & 0 & -10.0 & -42.0 & 0 & n/a \\
Z+HF(II) (\%) & 0 & 0 & -6.0 & -19.0 & 0 & n/a \\
Mistag (\%)& 22.0 & 15.0& 17.0 & 17.0 & 17.0 & n/a\\
Top I (\%)&13.5&20.0& 12.0 & 12.0 & 20.0& n/a\\
Top II (\%)&n/a& n/a &-2.0 & -3.0 & n/a & n/a\\
QCD (\%)&17.0&20.0&-10.0 & -44.0 & -50.0 & n/a\\
Diboson I (\%)&16.0&25.0&12.0 & 12.0 & 20.0 & 11.0\\
Diboson I (\%)& n/a & n/a &-5.0 & -10.0 & n/a & n/a\\
Others (\%)& n/a&n/a&n/a&n/a& n/a & -(12.0-18.0)\\
\end{tabular}
\end{ruledtabular}
\end{table}

\subsubsection{\DZ Systematics}

The dominant systematic uncertainties for \DZ analyses are shown in
Table~\ref{tab:dzsyst1}. The \hbb~analyses have an uncertainty on the
$b$-tagging rate of 5-7\% per tagged jet.  These analyses also have an
uncertainty on the jet measurement and acceptance of 6-9\% (jet
identification or jet ID, energy scale, and jet smearing).  For the
\hww~and \www, the largest uncertainties are associated with lepton
measurement and acceptance. These values range from 3-6\% depending on
the final state.  The largest contributing factor to all analyses is
the uncertainty on the cross sections for inherent background at
6-15\%. These systematics are assumed to apply to both signal and
background processes.  All systematic uncertainties arising from the
same source are taken to be correlated between signal and
background, as detailed in Table~\ref{tab:dzsyst1}.

\begin{table}
\caption{\label{tab:dzsyst1}List of leading correlated systematic
uncertainties for the \DZ analyses. The values for the systematic
uncertainties are the same for the \ZH~and \lmet~channels. All
uncertainties with a common origin are considered 100\% correlated
across channels. The correlated systematic uncertainty on the
background cross section ($\sigma$) is itself subdivided according
to the different background processes in each analysis.}
\begin{ruledtabular}
\begin{tabular}{lccc}
\\
Source &$WH\rightarrow e\nu b\bar{b}$ DT(ST) &$WH\rightarrow \mu\nu b\bar{b}$ DT(ST)& $H\rightarrow W^+ W^-$, \www \\\hline
Luminosity (\%)& 6.5& 6.5& 6.5\\
Jet Energy Scale (\%)& 4.0& 5.0& 3.0\\
Jet ID (\%)& 6.8& 6.8& 0\\
Electron ID (\%)& 6.6& 0&2.3\\
Muon ID (\%)& 0& 4.9&7.7\\
$b$-Jet Tagging (\%)& 8.5(5.0)& 8.5(5.0)&0\\
Background $\sigma$ (\%)&6.0-19.0&6.0-19.0&6.0-19.0\\\hline
\\
Source & \ZH~DT(ST) &\zhe&\zhm\\\hline
Luminosity (\%)& 6.5& 6.5& 6.5\\
Jet Energy Scale (\%)& 6.0&7.0&2.0\\
Jet ID (\%)& 7.1&7.0&5.0\\
Electron ID (\%)& 0&8.0&0\\
Muon ID (\%)& 0&0&12.0\\
$b$-Jet Tagging (\%)& 9.6(6.7)&12.0&22.0\\
Background $\sigma$ (\%)&6.0-19.0&6.0-19.0&6.0-19.0\\
\end{tabular}
\end{ruledtabular}
\end{table}

\section{Combined Results} 

\begin{figure}[ht]
\begin{centering}
\includegraphics[width=12.0cm]{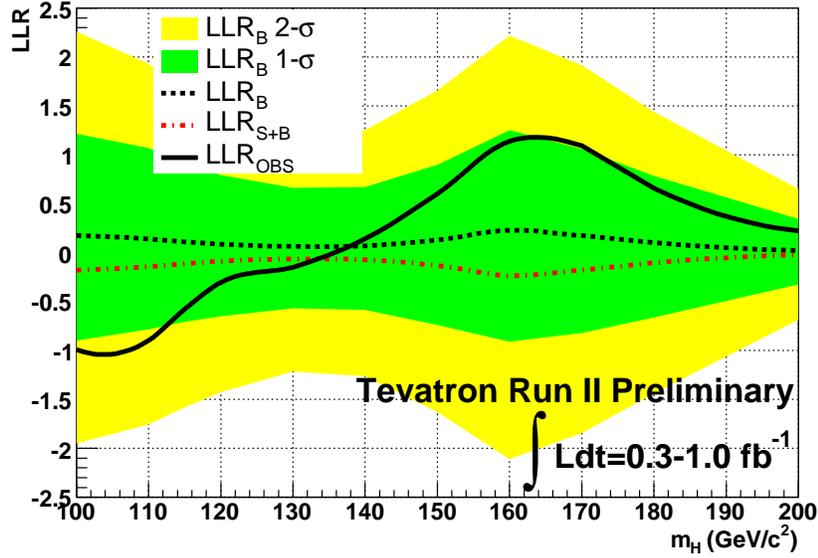}
\caption{
\label{fig:comboLLR}
Log-likelihood ratio distribution for the combined CDF and \DZ
analyses.  Shown in the plot are the LLR$_b$ (background-only
hypothesis), LLR$_{s+b}$ (signal+background hypothesis), LLR$_{obs}$
(observed LLR value), and the 1-$\sigma$ and 2-$\sigma$ bands for the
LLR$_b$ distribution.}
\end{centering}
\end{figure}

Using the combination procedures outlined above, we extract limits on
SM Higgs boson production $\sigma \times BR*(H\rightarrow X)$ in
\pp~collisions at $\sqrt{s}=1.96$~TeV. Figure~\ref{fig:comboLLR}
displays the log-likelihood ratio distributions for the combined
analyses as a function of $m_{H}$. Included are the results for the
background-only hypothesis (LLR$_{b}$), the signal+background
hypothesis (LLR$_{s+b}$), and the observed data (LLR$_{obs}$).  The
shaded bands represent the 1 and 2 standard deviation ($\sigma$)
departures for LLR$_{b}$. These distributions can be interpreted as
follows:

\begin{itemize}
\item The separation between LLR$_{b}$ and LLR$_{s+b}$ provides a
measure of the overall power of the search.  This is the ability of
the analysis to discriminate between the $s+b$ and $b-$only hypotheses.

\item The width of the LLR$_{b}$ distribution (shown here as 1 and 2
standard deviation bands) provides an estimate of how sensitive the
analysis is to a signal-like fluctuation in data, taking account of
the presence of systematic uncertainties.  For example, when a
1-$\sigma$ background fluctuation is large compared to the signal
expectation, the analysis sensitivity is thereby limited.

\item The value of LLR$_{obs}$ relative to LLR$_{s+b}$ and LLR$_{b}$
indicates whether the data distribution appears to be more signal-like
or background-like.  As noted above, the significance of any departures
of LLR$_{obs}$ from LLR$_{b}$ can be evaluated by the width of the
LLR$_{b}$ distribution.

\end{itemize}

\begin{figure}[ht]
\begin{centering}
\includegraphics[width=12.0cm]{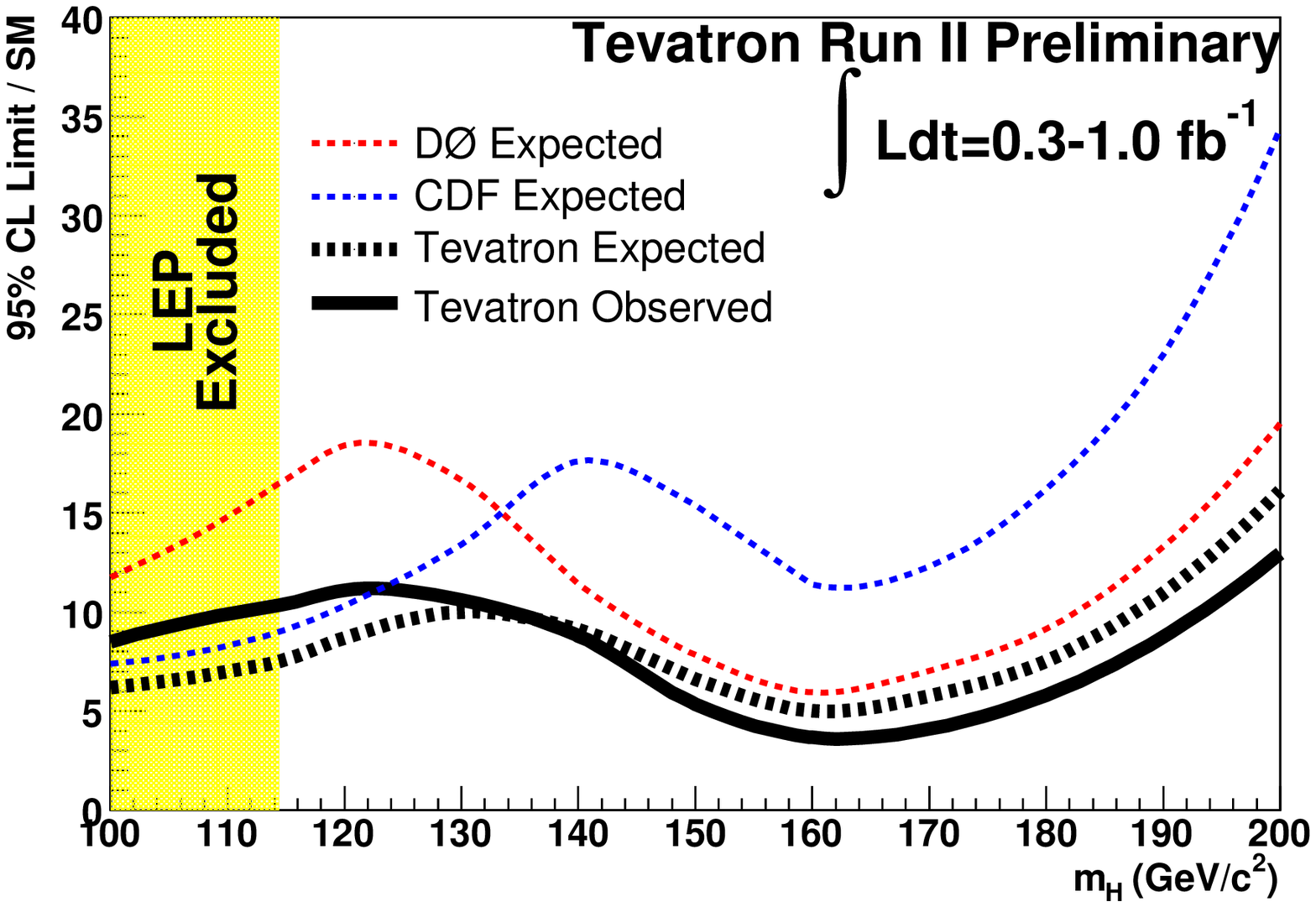}
\caption{
\label{fig:comboRatio}
Expected and observed 95\% CL cross section ratios for the
combined CDF and \DZ analyses. Also shown are the expected 95\% CL
ratios for the CDF and \DZ experiments alone.}
\end{centering}
\end{figure}

To facilitate model transparency and to accommodate analyses with
different degrees of sensitivity, we present our results in terms of
the ratio of limits set to the SM cross sections as a function of
Higgs mass.  A value of $<1$ would indicate a Higgs mass excluded at
95\% CL. The expected and observed 95\% upper limit ratios to the
SM cross section for the combined CDF and \DZ analyses are shown in
Figure~\ref{fig:comboRatio}.  The observed and expected limit ratios
are listed for selected Higgs masses in Table~\ref{tab:ratios}, with
observed(expected) values of 10.4(7.6) at $m_{H}=115$~GeV/c$^2$ and
3.8(5.0) at $m_{H}=160$~GeV/c$^2$.

These results represent an improvement in search sensitivity over
those obtained for individual experiments, which have set
observed(expected) limit ratios of 16.3(16.7) for \DZ and 12.8(9.1)
for CDF at $m_{H}=115$~GeV/c$^2$ and of 4.3(5.9) for \DZ and
11.4(10.2) for CDF at $m_{H}=160$~GeV/c$^2$.



\clearpage

\begin{table}
\caption{\label{tab:ratios}Expected and observed 95\% CL
cross section ratios for the combined CDF and \DZ analyses.}
\begin{ruledtabular}
\begin{tabular}{lccccccc}
\\
& 100 GeV/c$^2$& 115 GeV/c$^2$& 120 GeV/c$^2$& 140 GeV/c$^2$& 160 GeV/c$^2$& 180 GeV/c$^2$& 200 GeV/c$^2$\\ \hline 
Expected & 6.2 &  7.6&  8.7& 9.3& 5.0& 7.5& 15.5\\
Observed & 8.5 & 10.4& 11.1& 8.8& 3.8& 6.1& 12.3\\
\end{tabular}
\end{ruledtabular}
\end{table}

\begin{thebibliography}{000}

\bibitem{DZhiggs}
\DZero Collaboration, Conference Note 5056, ``Limits on Standard Model Higgs Boson Production''

\bibitem{CDFhiggs} CDF Collaboration, "Combined Upper Limit on
Standard Model Higgs Boson Production at CDF for Summer 2006,''
CDF/ANAL/EXOTIC/PUBLIC/8403

\bibitem{pythia} 
T.~Sjostrand, L.~Lonnblad and S.~Mrenna,
   ``PYTHIA 6.2: Physics and manual,''
  [arXiv:hep-ph/0108264]

\bibitem{cteq} 
H.~L.~Lai {\it et al.}, \textit{Improved Parton
Distributions from Global Analysis of Recent Deep Inelastic Scattering
and Inclusive Jet Data}, Phys. Rev D \textbf{55} (1997) 1280

\bibitem{nnlo1} 
S.~Catani, D.~de Florian, M.~Grazzini and P.~Nason,
   ``Soft-gluon resummation for Higgs boson production at hadron colliders,''
  JHEP {\bf 0307}, 028 (2003),
  [arXiv:hep-ph/0306211]

\bibitem{nnlo2} 
K.~A.~Assamagan {\it et al.}  [Higgs Working Group Collaboration],
   ``The Higgs working group: Summary report 2003,''
  [arXiv:hep-ph/0406152]

\bibitem{hdecay}
A.~Djouadi, J.~Kalinowski and M.~Spira,
   ``HDECAY: A program for Higgs boson decays in the standard model and its
   supersymmetric extension,''
  Comput.\ Phys.\ Commun.\  {\bf 108}, 56 (1998)
  [arXiv:hep-ph/9704448]

\bibitem{alpgen}
M.~L.~Mangano, M.~Moretti, F.~Piccinini, R.~Pittau and A.~D.~Polosa,
   ``ALPGEN, a generator for hard multiparton processes in hadronic
   collisions,''
  JHEP {\bf 0307}, 001 (2003)
  [arXiv:hep-ph/0206293]

\bibitem{herwig} 
G.~Corcella {\it et al.},
   ``HERWIG 6: An event generator for hadron emission reactions with
   interfering gluons (including supersymmetric processes),''
  JHEP {\bf 0101}, 010 (2001)
  [arXiv:hep-ph/0011363]

\bibitem{comphep}
A.~Pukhov {\it et al.},
   ``CompHEP: A package for evaluation of Feynman diagrams and integration  over
   multi-particle phase space. User's manual for version 33,''
  [arXiv:hep-ph/9908288]

\bibitem{mcfm}
 J.~Campbell and R.~K.~Ellis,
 ``Next-to-leading
order corrections to W + 2jet and Z + 2jet production at
 hadron
colliders,''
 Phys.\ Rev.\ D {\bf 65}, 113007 (2002),
[arXiv:hep-ph/0202176]
\bibitem{cdfWH} CDF Collaboration, "Search for Higgs Boson Production
in Association with W Boson with 1~\ifb," CDF/ANAL/EXOTIC/PUBLIC/8390

\bibitem{cdfZH}  CDF 1~\ifb ZH result.
\bibitem{cdfZHll} CDF 1~\ifb ZH result in dilepton channel.

\bibitem{cdfHWW} CDF Collaboration, ``Search for the Standard Model
Higgs Boson in the $gg\rightarrow H \rightarrow WW^{*}$ Dilepton
Channel with 360~\ipb,'' CDF/ANAL/EXOTIC/PUBLIC/7893

\bibitem{dzWHl} \DZ Collaboration, ``Search for WH Production at
$\sqrt{s}=1.96$~TeV,'' \DZ Conference Note 5054
\bibitem{dzZHv} \DZ Collaboration, ``A Search for the Standard Mondel
Higgs boson using the \ZH~channel in \pp~Collisions at
$\sqrt{s}=1.96$~TeV,'' submitted to Phys. Rev. Lett., [arXiv:hep-ex/0607022]
\bibitem{dzZHll} \DZ Collaboration, ``A Search for \ZHll Production at
  \DZ in \pp Collisions at $\sqrt{s}=1.96$~TeV,'' \DZ Conference Note 5186
\bibitem{dzWWW} \DZ Collaboration, ``Search for associated Higgs boson
production $WH\rightarrow WWW^* \rightarrow \ell^\pm \nu
\ell^{\prime\pm} \nu^\prime +X$ in $p\bar{p}$ collisions at
$\sqrt{s}=1.96$~TeV,'' submitted to Phys. Rev. Lett.,
[arXiv:hep-ex/0607032]
\bibitem{dzHWWee} \DZ Collaboration, ``Search for the Higgs boson in
$H \rightarrow W W^* \rightarrow l^+ l^- (ee, e \mu)$ decays with
950~\ipb~at \DZ in Run II,'' \DZ Conference Note 5063
\bibitem{dzHWWmm} \DZ Collaboration, ``Search for the Higgs boson in
$H \rightarrow W W^* \rightarrow \mu\mu$ decays with 930~\ipb~at \DZ
in Run II,'' \DZ Conference Note 5194


\end{thebibliography}
\end{document}